\documentclass[aps,pra,reprint,a4paper,superscriptaddress,numbers,twocolumn,longbibliography,showpacs,showkeys]{revtex4-1}
\usepackage[pdftex]{hyperref,color,graphicx}
\usepackage{amsfonts,amssymb,amsmath}
\usepackage[separate-uncertainty=false]{siunitx}
\usepackage[english]{babel}
\usepackage[T1]{fontenc}
\usepackage{comment}
\usepackage{xspace}
\usepackage{dcolumn}
\usepackage{bm}
\usepackage{braket}
\usepackage{verbatim,ulem}
\usepackage{mathtools}
\usepackage{tikz}

\newcommand{\red}[1]{\textcolor{black}{#1}}

\usetikzlibrary{decorations.pathmorphing}

\DeclareMathOperator{\tr}{tr}
\newcommand{\bbrapprox}{\overset{\makebox[0pt]{\mbox{\normalfont\scriptsize BBR}}}{\approx}}

\begin{document}

\title{ \red{Asymmetric many-body loss in a bosonic double well} }

\author{Zakari Denis}
\affiliation{Dipartimento di Scienze Matematiche, Fisiche e Informatiche, 
Campus Universitario, Parco Area delle Scienze n.\ 7/a, Universit\`a di Parma, 
43124 Parma, Italy}
\affiliation{Université Paris-Sud (Université Paris-Saclay), 91405 Orsay, 
France}

\author{Antonio Tiene}
\affiliation{Dipartimento di Fisica e Astronomia ``Galileo Galilei'', 
Universit\`a di Padova, Via Marzolo 8, 35131 Padova, Italy}

\author{Luca Salasnich}
\affiliation{Dipartimento di Fisica e Astronomia ``Galileo Galilei'' and 
CNISM,  Universit\`a di Padova, Via Marzolo 8, 35131 Padova, Italy}
\affiliation{Istituto Nazionale di Ottica (INO) del Consiglio Nazionale delle 
Ricerche (CNR), Via Nello Carrara 1, 50019 Sesto Fiorentino, Italy}

\author{Sandro Wimberger}
\affiliation{Dipartimento di Scienze Matematiche, Fisiche e Informatiche, 
Campus Universitario, Parco Area delle Scienze n.\ 7/a, Universit\`a di Parma, 
43124 Parma, Italy}
\affiliation{INFN, Sezione di Milano Bicocca, Gruppo Collegato di Parma, 
43124 Parma, Italy}

\begin{abstract}
A Bose gas in a double well is investigated in the presence of single-particle, two-body and three-body asymmetric loss.
The loss induces an interesting decay behavior of the total population as well as a possibility to control the dynamics of the system. In the noninteracting limit with asymmetric single-body dissipation, the dynamics of the populations can be obtained analytically. The general many-body problem requires, however, an adequate approximation. We use a mean-field approximation and the Bogoliubov back-reaction beyond mean-field truncation, which we extend up to three-body loss. Both methods are compared with exact many-body Monte-Carlo simulations. 
\end{abstract} 

\pacs{
03.75.Gg, 
05.60.-k 
}
 
\keywords{Open quantum systems, Bose-Einstein condensates, Ultracold atoms}

\date{\today}

\maketitle

\section{Introduction}
\label{sec-intro}

Open many-body quantum systems offer a huge playground of emergent phenomena. Of particular interest is a situation in which the time-scales of various processes are matched such as to arrive at 'resonant' dynamics of the system. A nice example is the famous phenomenon of stochastic resonance \cite{Review1, Review2}, which can be exported also onto the quantum level, see e.g. \cite{Review2, Witthaut2009}. Interestingly, local loss can also enhance the global coherence properties of a many-body quantum system, see e.g. \cite{Castin2008, Witthaut2008, Witthaut2008JPB, Witthaut2009, Zoller, Cirac}.

\begin{figure}[tb!]
\begin{center}
\includegraphics[width=0.9\columnwidth]{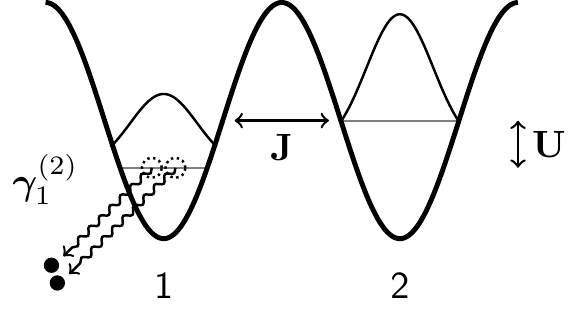}
\end{center}
\caption{\label{fig:1}%
	Schematic illustration of a dissipative double well. Tunneling, characterised by the energy scale $J$, is allowed between the wells. Interactions are denoted by the energy scale $U$, and an exemplary two-body loss process from the left well (with index 1) is shown by the wiggly lines.}
\end{figure}

In order to keep the discussion on a \red{simple} level, we address here mostly the problem of a quantum mechanical double well filled with interacting bosons, see Fig. \ref{fig:1}. This system is well studied experimentally with Bose-Einstein condensates (BEC). \red{In such experiments, typically a BEC of rubidium 87 atoms is loaded into a double-well structure formed by the superposition of an harmonic trap and a periodic light potential. The potential and the number of atoms are well controlled, the latter almost down to shot-noise precision, see e.g. \cite{Albiez2005, Oberthaler2010}.} We allow for \red{asymmetric loss in the two wells}, which can be of single-\red{body}, two-body or even three-body nature, modelling experimentally relevant decay processes, which are either induced by external out couplings \cite{Gericke2008, Wuertz2009, Barontini2013, Bloch2011}, or by two- or three-body scattering processes \cite{Wieman2000}. For instance, three-body recombination in an optical lattice leads to decay into unbound states and thus to loss from the lattice \cite{Kraemer2006}. 

Our main findings are {\it (i)} a total decay of an initially asymmetric population of the wells in form of a 'staircase' which is robust with respect to interactions and the precise form of the loss, and  {\it (ii)} a way to control dynamically the evolution of the system \red{by an appropriate choice of the loss}. Both effects are induced by a time-scale matching between the two-mode oscillation frequency, the loss rates, and possible amendments arising from the interactions. We start discussing these effects on a mean-field level with exact results in the non-interacting case. Extensions beyond mean-field and many-body simulations corroborate the stability of the findings. This robustness should allow for an experimental observation of our predictions with state-of-the-art apparatuses.

\red{The paper is organised as follows: Sec. \ref{sec-2} presents the two-site many-body Hamiltonian and the master equation for the open system's evolution. Sec. \ref{sec-3} reports the analytical solution for the linear non-interacting problem, which is easily extended in appendix \ref{sec-appendix-a} to an arbitrarily large one-dimensional system. The effect of nonlinear contributions, both by the atom-atom interactions as well as the many-body loss, on the observed decay is addressed in Sec. \ref{sec-4} in mean-field approximation. Sec. \ref{sec-5} compares the quality of our mean-field results with a beyond mean-field expansion and exact quantum trajectory simulations. Conclusions are presented in Sec. \ref{sec-con}.}

\section{Dissipative double well system}
\label{sec-2}

Without the loss, our system is described by the following Hamiltonian in second quantisation
\begin{equation}
\hat{H} = -J(\hat{a}^\dagger_1\hat{a}^{\mathstrut}_2+\hat{a}^\dagger_2\hat{a}^{\mathstrut}_1)+\frac{U}{2}(\hat{a}^\dagger_1\hat{a}^\dagger_1\hat{a}^{\mathstrut}_1\hat{a}^{\mathstrut}_1+\hat{a}^\dagger_2\hat{a}^\dagger_2\hat{a}^{\mathstrut}_2\hat{a}^{\mathstrut}_2) \,.
\label{eq:1}
\end{equation}
$\hat{a}^{\mathstrut}_i$ and $\hat{a}^\dagger_i$ are the annihilation and creation operators of bosons for the two modes $i=1,2$, respectively. \red{$\hbar$ is set to 1, thus measuring energies in frequency units.} Including the particle losses, we start from a master equation including the imaginary shift of the Hamiltonian as well as the quantum jumps from losing particles:
\begin{equation}
\dot{\hat{\rho}} = -i(\hat{H}_{\textrm{eff}}\hat{\rho}-\hat{\rho}\hat{H}^{\dagger}_{\textrm{eff}})+\sum_\alpha\sum_{\ell=1,2}\gamma^{(\alpha)}_\ell\hat{a}^{\mathstrut\alpha}_\ell\hat{\rho}\hat{a}^{\dagger\alpha}_\ell \,.
\label{eq:2}
\end{equation}
The effective Hamiltonian reads
\begin{equation}
\hat{H}_{\textrm{eff}}=\hat{H}-i\sum_\alpha\sum_{\ell=1,2}\frac{\gamma^{(\alpha)}_\ell}{2}\hat{a}^{\dagger \alpha}_\ell\hat{a}^{\mathstrut\alpha}_\ell \,
\label{eq:3}
\end{equation}
where the index $\alpha$ stands for 1, 2, or 3, respectively, describing any of the used forms of $\alpha$-body loss.

We will start with a mean-field method introduced in \cite{Anglin2001} and heavily used previously for bosons in lattice structures subject to decay \cite{Witthaut2011, Trimborn2011, Kordas2015}. The method has the advantage that the results are easily extendible to any number of sites $M$ (the number of terms scale just like $M^2$ independently of the particle number) and at least in the non-interacting case we can come up with analytical expressions in the presence of single-body loss.

The \red{time-dependent} populations of the two modes are given by \red{the diagonal elements of} the single-particle reduced density matrix (SPDM) \smash{$\sigma_{j,k} = \tr\lbrace\hat{a}^\dagger_j\hat{a}^{\mathstrut}_k\hat{\rho}\rbrace\equiv\langle\hat{a}^\dagger_j\hat{a}^{\mathstrut}_k\rangle$} ($j,k=1,2$). In the mean-field approximation and for $\alpha$--body losses, its four elements obey the following equations of motion
\begin{align}
i\frac{d}{dt}\sigma_{j,k}&=tr(\hat{a}^\dagger_j \hat{a}_k[\hat{H},\hat{\rho}]+i\hat{a}^\dagger_j \hat{a}_k\mathcal{L}_{\textrm{tot}}\hat{\rho})\nonumber\\
=&-J(\sigma_{j,k+1}+\sigma_{j,k-1}-\sigma_{j+1,k}-\sigma_{j-1,k})\label{eq:4}\\
&+U\sigma_{j,k}(\sigma_{k,k}-\sigma_{j,j})+i\textstyle\sum_\alpha\tr\lbrace\hat{a}^\dagger_j \hat{a}_k\mathcal{L}^{(\alpha)}\hat{\rho}\rbrace\nonumber \, .
\end{align}
We used the definitions
\begin{align}
\mathcal{L}^{(\alpha)}\hat{\rho}=-\sum_{\mathclap{\ell=1,2}}\frac{\gamma^{(\alpha)}_\ell}{2}(\hat{a}^{\dagger\alpha}_\ell\hat{a}^{\mathstrut\alpha}_\ell\hat{\rho}+\hat{\rho}\hat{a}^{\dagger\alpha}_\ell\hat{a}^{\mathstrut\alpha}_\ell-2\hat{a}^{\mathstrut\alpha}_\ell\hat{\rho}\hat{a}^{\dagger\alpha}_\ell)
\label{eq:5}\\
\tr\lbrace\hat{a}^\dagger_j\hat{a}_k\mathcal{L}^{(\alpha)}\hat{\rho}\rbrace\overset{\mathclap{\textrm{MF}}}{=}-\sum_{\mathclap{\ell=1,2}}\frac{\gamma^{(\alpha)}_\ell}{2}\sigma_{j,k}(\delta_{j\ell}+\delta_{k\ell})f_\alpha(n_\ell)
\label{eq:6}
\end{align}
with $f_\alpha$ a polynomial of degree $(\alpha-1)$:
\begin{align}
&f_1(n_\ell)=1\,, \qquad f_2(n_\ell)=2(n_\ell-1)\,,\label{eq:7}\\
&f_3(n_\ell)=3(n^2_\ell-3n_\ell+2)\nonumber \, .
\end{align}

From Eq.(\ref{eq:7}), one observes that, for $N\gg\alpha$, $\alpha$-body loss terms scale as $N^{\alpha-1}$ with respect to those of single-body loss. Therefore, we henceforth rescale the $\alpha$-body dissipation rates by a factor $N_0^{\alpha-1}$ for a better comparison between the various losses. \red{From the form of the equations, we see that loss with $\alpha > 1$ induces a time-evolution, which depends on the local densities in a nonlinear manner.}

\section{Single-body decay}
\label{sec-3}

\begin{figure}[tb!]
\begin{center}
\includegraphics[width=\columnwidth]{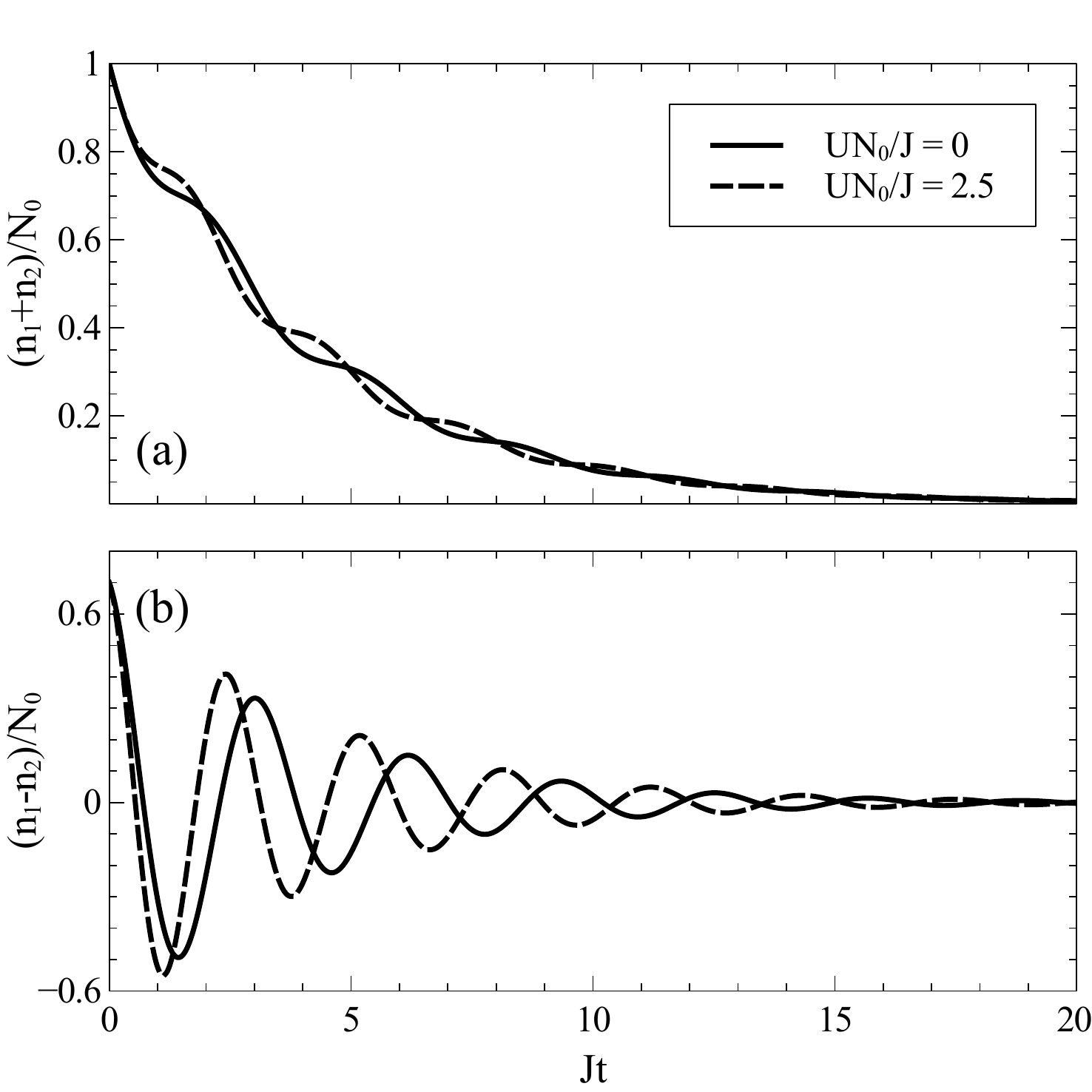}
\caption{\label{fig:2}
Mean-field evolution of \red{the total} $N$ (a) and \red{the relative population} $\Delta n$ (b) without (solid-lines) and with (dashed lines) interactions with single-body losses, for the parameters $\Delta n(0)/N_0=0.7$, \smash{$\gamma_1^{(1)}/J=0.5$}, \smash{$\gamma_2^{(1)}/J=0$}.
}
\end{center}
\end{figure}

The \red{time evolution of the} SPDM has a particularly simple form for single-body ($\alpha=1$) loss and without interactions. Its four matrix elements obey the following \red{linear} equations of motion:
\begin{align}
i\frac{d}{dt}\sigma_{j,k}=&-J(\sigma_{j,k+1}+\sigma_{j,k-1}-\sigma_{j+1,k}-\sigma_{j-1,k})\label{eq:8}\\
&+\frac{\gamma_j^{(1)}+\gamma_k^{(1)}}{2}\sigma_{j,k}\nonumber \,.
\end{align}
Let us denote by $\vec{\sigma}(t)$ the vector consisting of the four matrix elements of the SPDM. Rearranging Eq. (\ref{eq:8}) under the form $\partial_t \vec{\sigma}(t)=\mathbf{A}\vec{\sigma}(t)$, the general solution is then obtained by integrating this equation expressed in the eigenbasis of the matrix $\mathbf{A}$. For any set of initial conditions, the solution can be put into the form $\vec{\sigma}(t)=\mathbf{U}(t)\vec{\sigma}(0)$, from which one derives the quantities of interest, for instance for asymmetric dissipation (\smash{$\gamma_1^{(1)}/J\neq 0$}, \smash{$\gamma_2^{(1)}/J=0$}):
\begin{align}
\frac{N(t)}{N_0}&=e^{-\frac{\gamma_1 t}{2}}\left(\frac{1-(\tfrac{x}{4})^2 \cosh(2J yt)}{1-(\tfrac{x}{4})^2}\right.\nonumber\\&\qquad\qquad\qquad\qquad\left.-\frac{\Delta n(0)}{N_0}\frac{(\tfrac{x}{4})\sinh(2J yt)}{y}\right)\label{eq:9}\\
\frac{\Delta n(t)}{N_0}&=e^{-\frac{\gamma_1 t}{2}}\left(-\frac{\tfrac{x}{4}\sinh(2J yt)}{y}+\frac{\Delta n(0)}{N_0}\cosh(2J yt)\right) \,,
\label{eq:10}
\end{align}
where \smash{$x\equiv \gamma_1^{(1)}/J$}, $y\equiv\sqrt{(x/4)^2-1}$, $N=n_1+n_2$ and $\Delta n=n_1-n_2$. These quantities exhibit two different regimes depending on whether $y\in\mathbb{R}$ (\smash{$\gamma_1^{(1)}/2>2J$}: monotonous decay) or $y\in i\mathbb{R}$ (\smash{$\gamma_1/2<2J$}: damped oscillations at the frequency $\omega=2J\lvert y\rvert$). 

In the latter case, the total population exhibits a step-like evolution, cf. Fig. \ref{fig:2}. \red{This can be understood by rewriting Eq. (\ref{eq:9}) as
\begin{align}
\frac{N(t)}{N_0}&=e^{-\frac{\gamma_1 t}{2}}\frac{1- C\sin(2J \lvert y\rvert t+\varphi_0)}{1-(x/4)^2}\label{eq:11}\,,
\end{align}
with
\begin{align}
C&=(x/4)^2\sqrt{1+(\tfrac{\Delta n(0)}{N_0})^2((\tfrac{4}{x})^2-1)}\label{eq:12}\,.
\end{align}
The equidistant steps become sharper the larger the initial imbalance is chosen, as the amplitude $C$ of the damped oscillations grows with that ratio. Moreover, one can observe that $C$ is zero for no dissipation or for a perfect match of the two time scales $\gamma_1/2 = 2J$ ($x/4 = 1$). In both cases the steps disappear. This behavior is interesting since it shows the relevance of the matching of time scales announced in the introduction. For a perfect match of them the current of bosons tunnelling from the non-dissipative well is resonant with that dissipated resulting in a uniform exponential decay in both wells. For mismatched time scales and $\gamma_1/2 < 2J$, the population of the leaky well is driven by tunnelling-induced oscillations and the dissipation is therefore quenched periodically when most of the population has tunnelled to the non-dissipative well.} This gives rise to the observed step-like plateau structure in $N(t)$ seen in Fig. \ref{fig:2} (a).

Our solution above can be straightforwardly extended to a higher number of modes as described in appendix A, which includes the explicit expression of the matrix $\mathbf{U}$ in the hereby considered double-well case as well.

The addition of the usual quartic interaction term from Eq. \eqref{eq:1} results in the known Josephson effect, increasing the frequency of the oscillations by a factor that scales as $2J\sqrt{1-UN/J}$ \cite{Raghavan1999}, where $U$ denotes the strength of the interactions. In the presence of dissipation, the density decays upon evolution so that this frequency shift is only observable before times of the order of \smash{$2/\gamma_1^{(1)}$}, resulting in a phase shift of later oscillations (cf. Fig. \ref{fig:2}).

\section{Interactions and many-body losses}
\label{sec-4}

In the presence of interactions and $\alpha$--body losses ($\alpha=1,2,3$), the four SPDM elements obey the full Eq. \eqref{eq:4} and can \red{only} be numerically computed upon time evolution.

\begin{figure}[tb!]
\begin{center}
\includegraphics[width=\columnwidth]{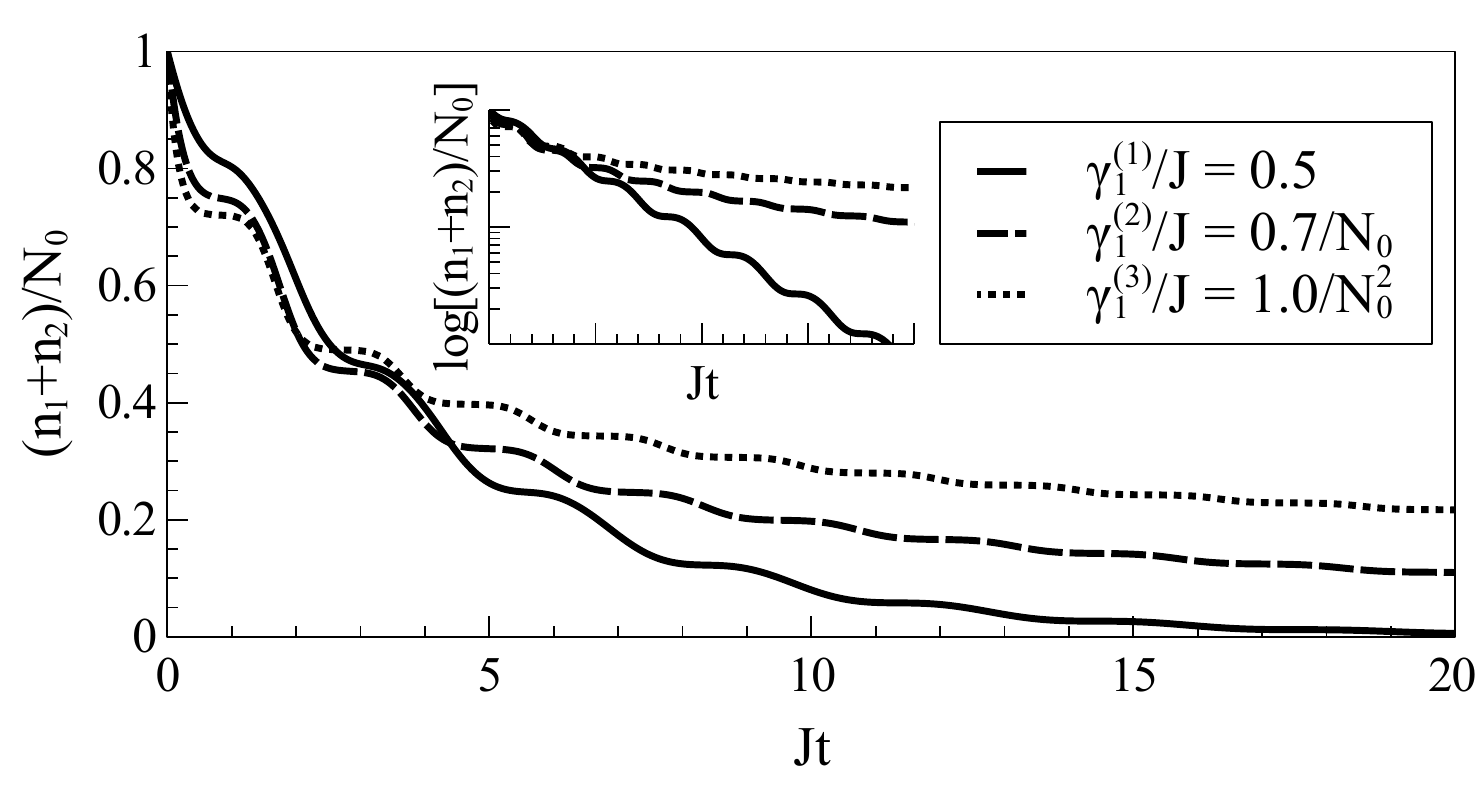}
\caption{\label{fig:3}%
Mean-field evolution of $N$ with interactions and $\alpha$--body losses in one site, see legend. The parameters are $\Delta n(0)/N_0=0.7$, $UN_0/J=10$. The inset shows the same data on a linear-logarithmic scale that allows one to appreciate the transient nonexponential decay in the cases of nonlinear two-body and three-body loss.
}
\end{center}
\end{figure}

Rich dynamics arise from the interplay of interactions and dissipation. Indeed, both effects add new time scales to the noninteracting and nondissipative characteristic tunneling time scale $2J$. While single-body losses shift the frequency of the population oscillations by a constant amount, interactions induce a dependence of the dynamics on the imbalance of the two wells. Many-body loss (i.e. with $\alpha > 1$) induce also a \red{nonlinear} dependence on the instantaneous population of the leaky well. \red{This} makes the decay nonexponential for early times. \red{Asymptotically in time, the decay becomes effectively exponential again} when the density is so weak that many-body corrections can be neglected. 

\red{Interaction-induced nonexponential tunnelling decay has been found in different contexts of BEC evolutions, see e.g. \cite{PRA2005, Schlagheck2007, PRA2010, Knoop2011}. This effect probes the role of interactions and can, in principle, be used to engineer the tunnelling decay, see  the example reported in reference \cite{Schlagheck2007}, where the compensation of an energy detuning from an internal resonance by the interaction-induced energy shift is exploited. For our double-well evolution, the difference between the studied $\alpha-$body decays is shown in 
Fig. \ref{fig:3}. The inset highlights there the nonexponential decay in a linear-logarithmic plot. While single-body loss decays perfectly exponentially, a two- or three-body loss dramatically deviates from the exponential form for short times, even if the loss rates are scaled according to the scaling rule found from Eq. \eqref{eq:7}. Our findings might be relevant for experiments since different contributions to decay could, in principle, be identified by their difference temporal behavior.}

\begin{figure}[tb!]
\begin{center}
\includegraphics[width=\columnwidth]{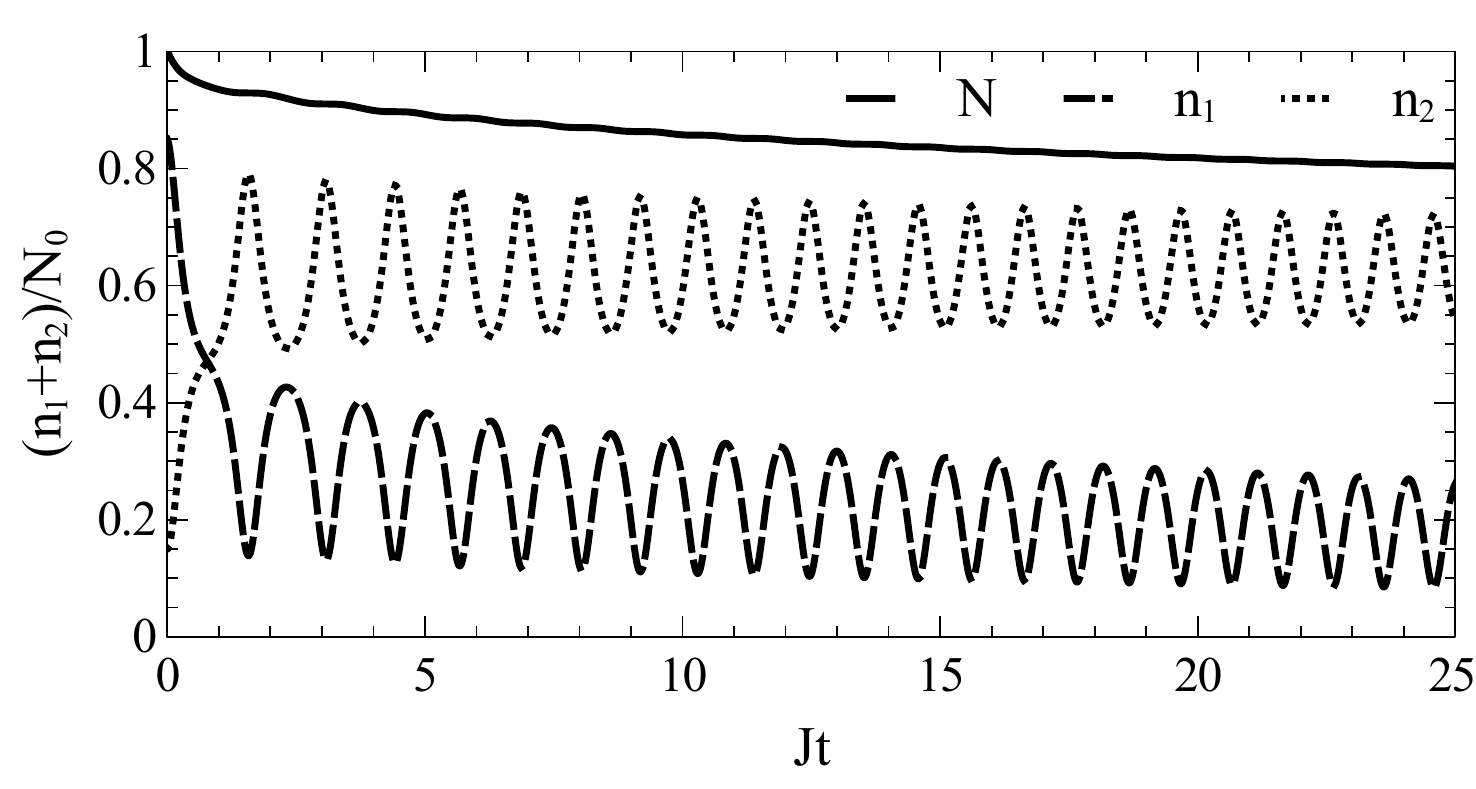}
\caption{\label{fig:4}%
Mean-field evolution of \red{the local populations} $n_1$, $n_2$ and \red{the total population} $N$ with interactions and three--body losses in one site, see legend. Parameter are $\Delta n(0)/N_0=0.7$, $UN_0/J=15$, \smash{$\gamma^{(3)}_1N^2_0/J=0.1$}.
}
\end{center}
\end{figure}

The density dependence of the many-body losses just discussed can be used for engineering the dynamics of the SPDM by tuning the initial imbalance, the interactions and the asymmetry of the dissipation. For example, it is possible to start from an imbalanced configuration and induce self-trapping in the \red{least} occupied site after a three--body loss driven inversion of the populations, as shown in the Fig. \ref{fig:4}. This way, the dissipation rate is naturally switched off \red{dynamically} by the depletion of the leaky site.

\begin{figure}[tb!]
\begin{center}
\includegraphics[width=\columnwidth]{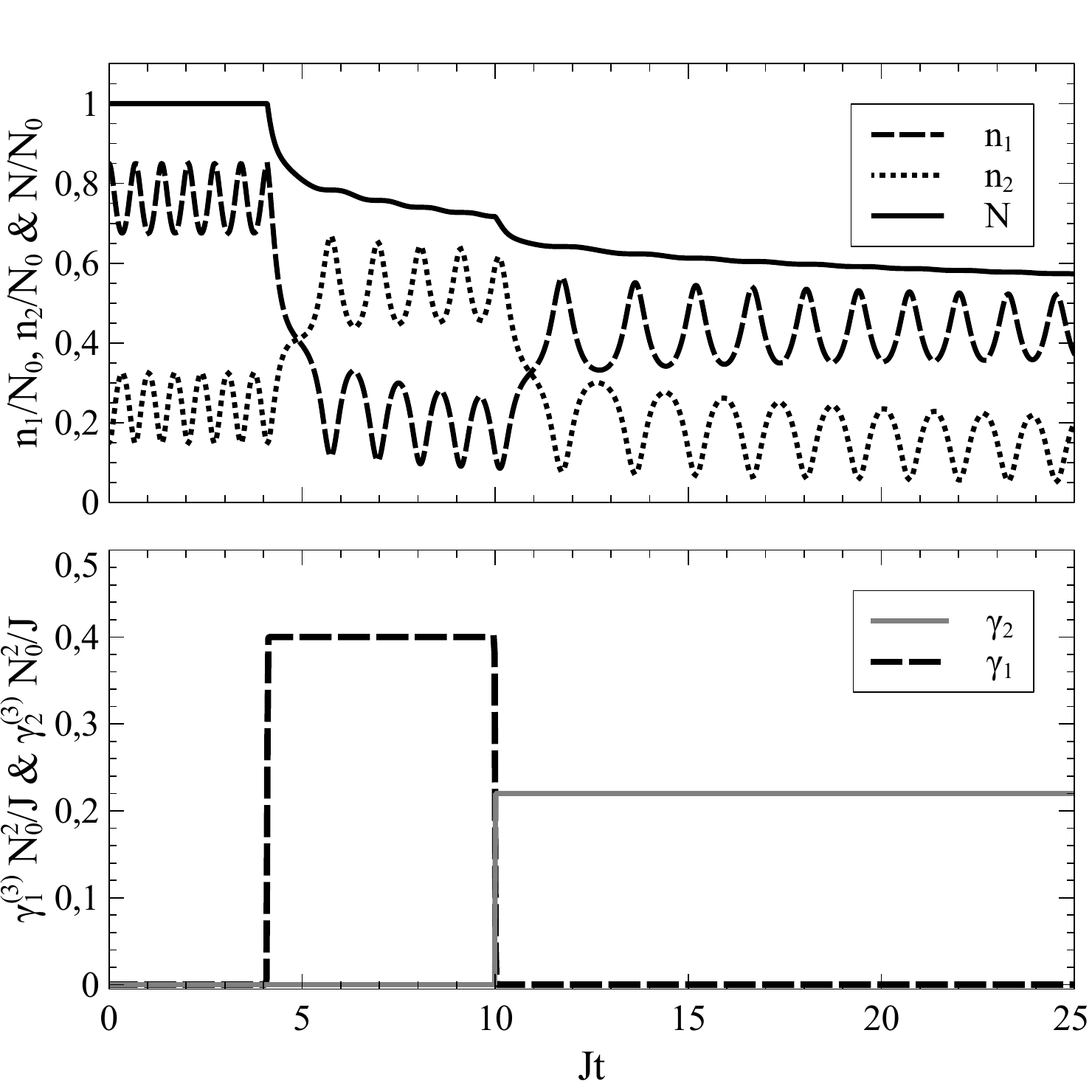}
\caption{\label{fig:4_1}
\red{
Upper panel: mean-field evolution of $n_1$, $n_2$ and $N$ with interactions and three--body losses in one site, for $\Delta n(0)/N_0=0.7$, $UN_0/J=18$. Lower panel: decay rates vs. time for the results in the upper panel, with \smash{$\gamma^{(3)}_1N^2_0/J=0.4$} for $4<Jt<10$, \smash{$\gamma^{(3)}_2N^2_0/J=0.22$} for $Jt>10$. }
}
\end{center}
\end{figure}

\red{In Fig. \ref{fig:4_1}, three-body dissipation is switched on in well $1$ in the time interval from $Jt \simeq 4$ to $Jt=10$. At  $Jt=10$ three-body loss is switched on in the opposite well, leading to a double inversion of the self-trapped population. Also here, the dissipation in well 1 is subsequently almost switched off by the depletion of the leaky site 2. Please note that the switching of the rates as shown in the lower panel of Fig. \ref{fig:4_1} does not need to be instantaneous, it should just be fast with respect to the tunnelling time scale $J^{-1}$.}

\red{We remark that the action of many-body dissipation in this kind of small Bose-Hubbard chains differs considerably from that of single-body loss. Indeed, whereas local single-body loss acts uniformly on the amplitude of the population oscillations by a factor $e^{-\gamma t/2}$ on both wells, as one sees from Eqs. (\ref{eq:9}) and (\ref{eq:10}), noting that $n_{1,2}(t) = (N(t)\pm\Delta n(t))/2$, the density dependence of many-body loss allows one to act separately on each site. As oscillations in the populations are in opposition of phase, one can act on the desired well when it reaches a population maximum and expect that the opposite well be protected to some extent by being at its population minimum, as done in the Fig. \ref{fig:4_1}.}

One important observation is that the steps in the evolution of the total population are a general feature independent of $\alpha$, resulting from the \red{mismatch} of the two main time scales: the tunnelling rate between imbalanced wells and the loss rate\red{, properly rescaled by $N_0^{\alpha-1}$, see Fig. \ref{fig:3} and the discussion after Eq. \eqref{eq:11}. }

\section{Beyond mean-field evolution}
\label{sec-5}

\red{ 
The previous section discussed the effect of nonlinear mean-field interactions and many-body loss. In the following, we confirm the validity of these findings by comparing with {\it (i)} the solutions of equations including a second-order corrections with respect to Eq. \eqref{eq:4}, and {\it (ii)} exact simulations based on quantum trajectories for the many-body master equation \eqref{eq:2}. Expected deviations are only found for very strong interactions and if loss is so fast that too low particle numbers are reached. In both cases, the mean-field expansions lose validity as further discussed in appendix \ref{sec:B4}. 
}

The hierarchy of equations of motion arising from interactions and many-body dissipation can be truncated to next order in the correlations by using the equations of motion satisfied by the covariances \smash{$\Delta_{jmkn} \equiv \langle \hat{a}_j^\dagger\hat{a}_m^{\mathstrut}\hat{a}_k^\dagger\hat{a}_n^{\mathstrut} \rangle - \langle \hat{a}_j^\dagger\hat{a}_m^{\mathstrut} \rangle\langle \hat{a}_k^\dagger\hat{a}_n^{\mathstrut} \rangle$} together with Eq.\,(\ref{eq:4}) and truncating higher order moments, please c.f. appendix\,\ref{sec-appendix-b} for details. This so-called Bogoliubov back-reaction (BBR) approximation takes approximately into account covariances with a relative error of order $1/N^2$ \cite{Anglin2001,Tikhonenkov2007}. It is thus only valid for close to pure Bose-Einstein condensates and for sufficiently large populations in all the wells.

\begin{figure}[tb!]
	\begin{center}
		\includegraphics[width=\columnwidth]{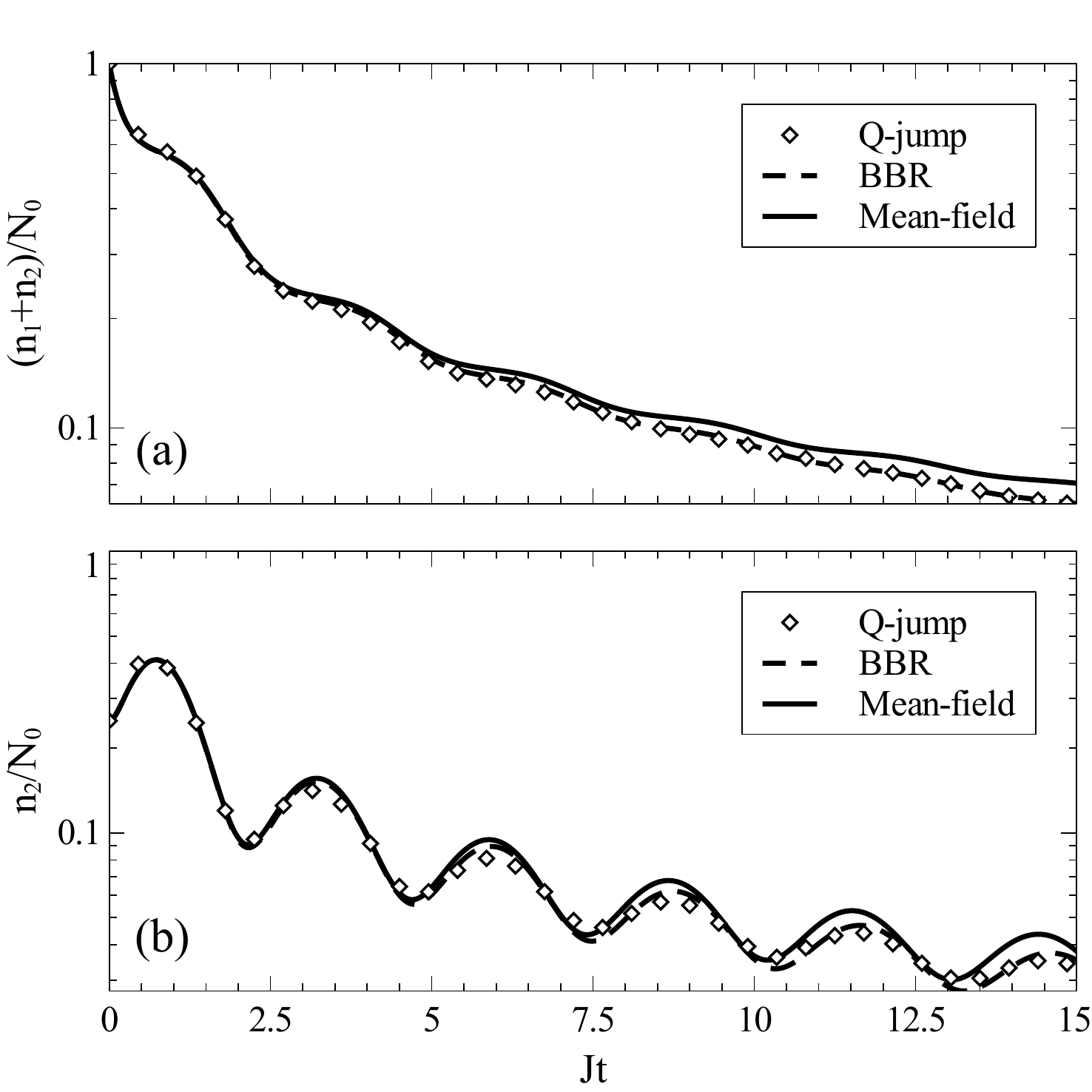}
		\caption{\label{fig:5}%
			Exact, BBR and mean-field evolution of $N$ (a) and $n_2$ (b) in the presence of interactions and 2-body dissipation. Parameters are $N_0=100$, $UN_0/J=5$ and \smash{$\gamma_1^{(2)}N_0/J=2$}. The numerically exact many-body data has been computed by using the quantum jump method averaging over 200 trajectories. Please note the logarithmic scale in both the ordinate axes.%
		}
	\end{center}
\end{figure}

The performance of both mean-field and BBR approximations is compared to exact many-body calculations in Fig. \ref{fig:5}, where BBR approximation is seen to provide a better description of the dynamics of the total population than mean-field in the presence of interactions and two-body dissipation. The nonexponential decay of the population is brought to light by the logarithmic scale. From this figure one also notes that, despite the low number of particles, the presence of two-body loss and the strength of the interactions, mean-field approximation still satisfactorily describes the system's evolution.

\begin{figure}[tb!]
	\begin{center}
		\includegraphics[width=\columnwidth]{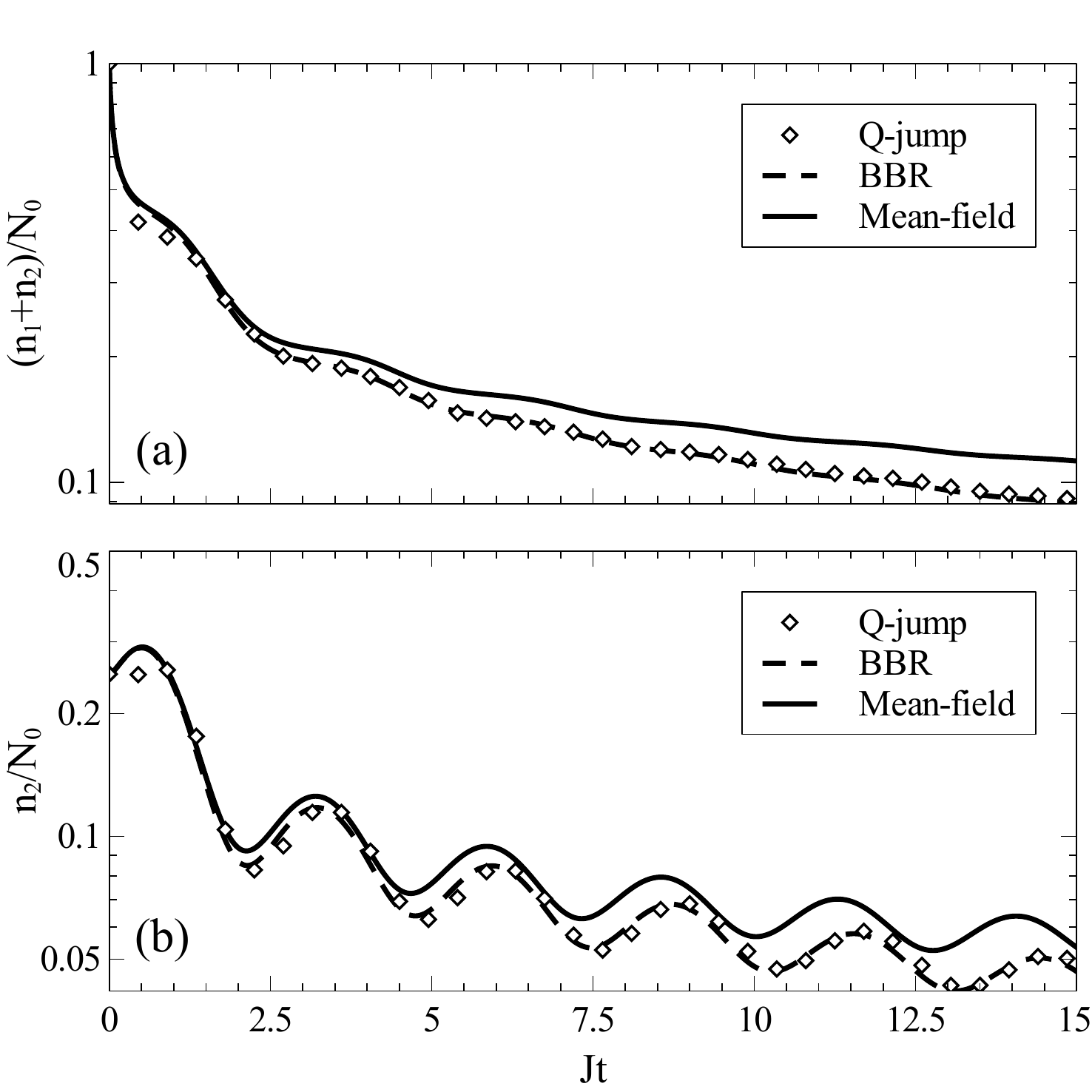}
		\caption{\label{fig:6}%
			Exact, BBR and mean-field evolution of $N_0$ (a) and $n_2$ (b) in the presence of interactions and 3-body dissipation. Parameters are $N_0=100$, $UN_0/J=5$ and \smash{$\gamma_1^{(3)}N_0^2/J=10$}. The numerically exact many-body data has been computed by using the quantum jump method averaging over 200 trajectories. %
		}
	\end{center}
\end{figure}

The performance of the BBR truncation with respect to mean-field in low populated situations becomes more striking in the case of strong three-body loss. This is shown in Fig.\,\ref{fig:6}, where mean-field systematically overestimate the populations of the double-well whereas BBR still performs satisfactorily.

\begin{figure}[tb!]
	\begin{center}
		\includegraphics[width=\columnwidth]{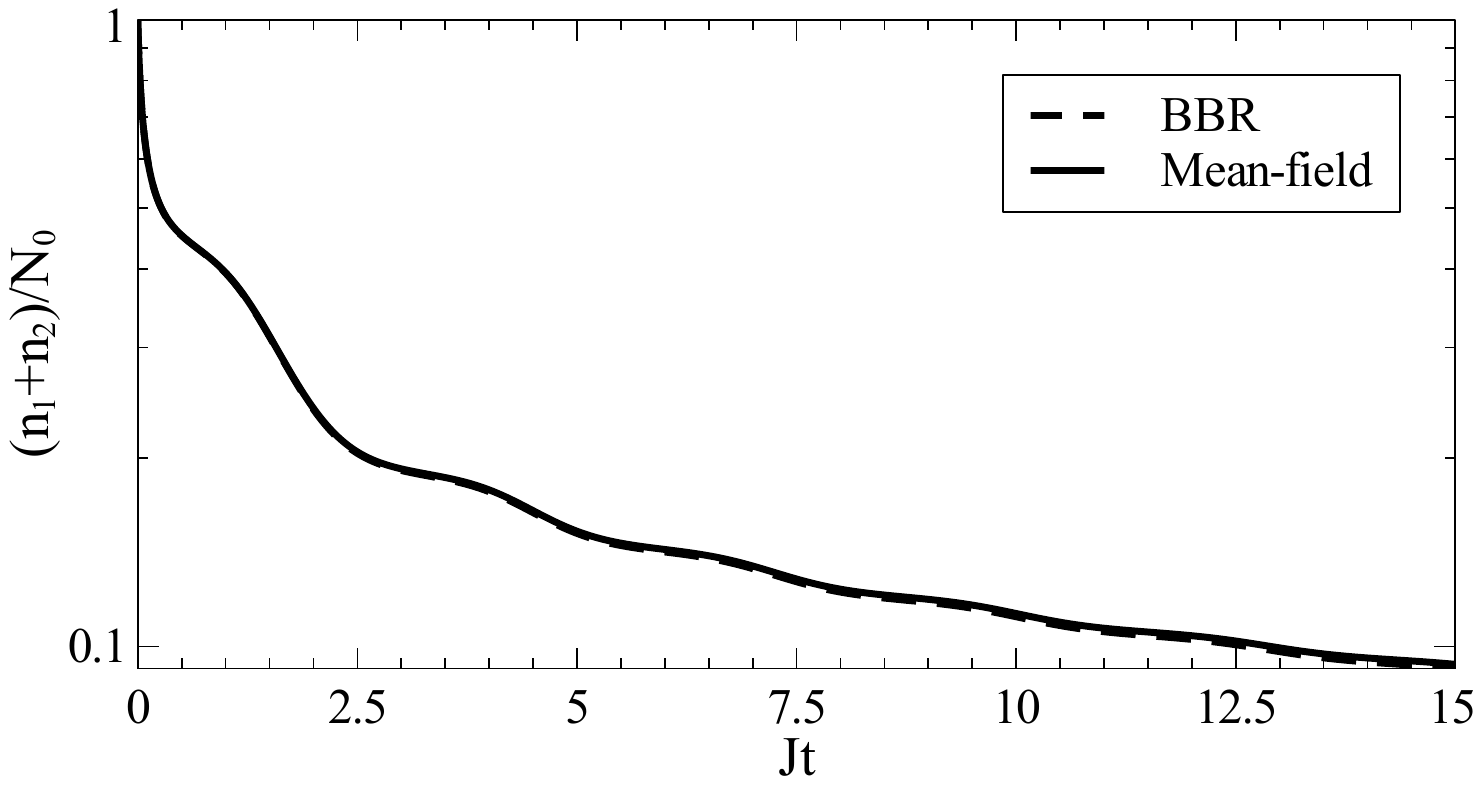}
		\caption{\label{fig:7}%
			BBR and mean-field evolution of $N$ in the presence of interactions and 2-body dissipation. Parameters are $N_0=1000$, $UN_0/J=5$ and \smash{$\gamma_1^{(3)}N_0^2/J=10$}. 
		}
	\end{center}
\end{figure}

Finally, as shown in Fig. \ref{fig:7}, BBR results for many-body dissipation converge asymptotically to those of mean-field as the ratio \red{$U/J$} is decreased to the regime of validity of the mean-field approximation \red{at fixed $UN_0/J=5$}, in this case by increasing the initial total population $N_0$.

\begin{figure}[tb!]
	\begin{center}
		\includegraphics[width=\columnwidth]{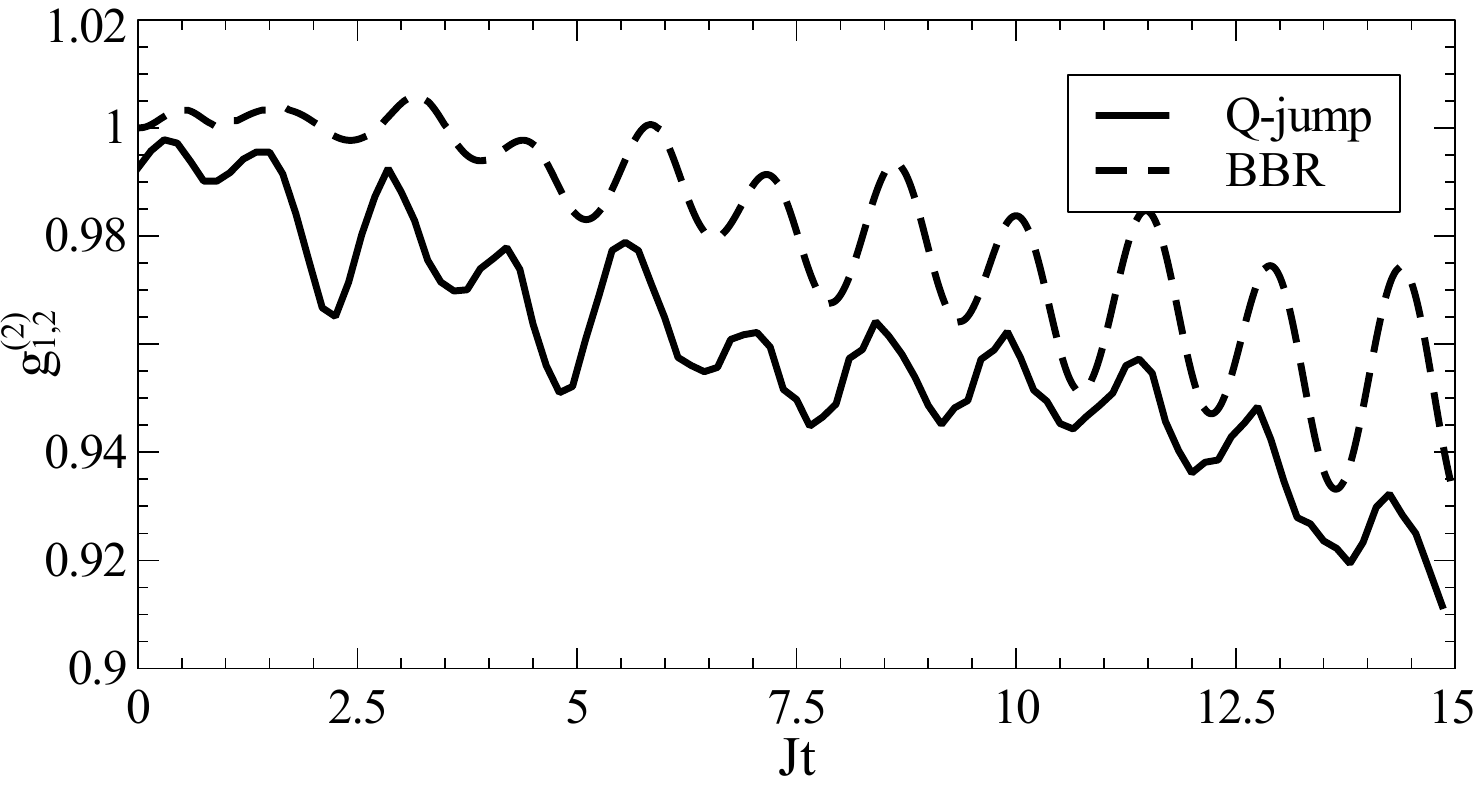}
		\caption{\label{fig:8}%
			Exact and BBR evolution of the $g^{(2)}_{1,2}$ correlation function. Parameters are $N_0=100$, $UN_0/J=5$ and \smash{$ \gamma_1^{(2)}N_0/J=2 $}. The numerically exact many-body data has been computed by using the quantum jump method averaging over 200 trajectories. Both curves essentially follow each other with a relative error of just a few percent. Mean-field would always give a value of one. The slow decay from one indicates the deterioration of the mean-field approximation in this case.%
		}
	\end{center}
\end{figure}

The BBR approximation allows one to compute two-point correlation functions, which, besides providing useful information about the fluctuations, constitute a good criterion for the validity of the approximation by evaluating whether the system remains in a coherent pure BEC state suitable for a mean-field study. As mean-field is no longer valid when variances cannot be neglected, a good criterion is provided by the density-density correlation function \red{\smash{$g^{(2)}_{1,2} \neq 1$}}, defined generally as follows
\begin{equation}
g^{(2)}_{j,k} \equiv \frac{\langle \hat{a}^\dagger_j\hat{a}^{\mathstrut}_j\hat{a}^\dagger_k\hat{a}^{\mathstrut}_k \rangle}{n_j n_k} \equiv 1+\frac{\Delta_{jjkk}}{n_j n_k} \,.
\label{eq:12}
\end{equation}
As shown in Fig. \ref{fig:8}, interactions and dissipation generally drive the system slowly away from a coherent BEC state and induce density-density correlations between the two wells. Nevertheless, the many-body evolution profile is well followed by the BBR approximation. \red{To summarize this section, both the mean-field and the BBR approximations are rather good as long as the populations remain large and self-trapping in one well due to a too large imbalance and/or too large interaction strength is avoided. This is corroborated by our many-body quantum jump simulations, which would become very tedious for \red{a larger number of bosons} however.}

\section{Conclusions}
\label{sec-con}

We have investigated and compared single-, two-, and three-body asymmetric loss in a two-mode system filled with bosons. The observed step-like decay in the system is observed qualitatively in all cases, provided that the initial state populations are imbalanced between the two wells. Loss can attack at both wells simultaneously, making a possible experimental implementation simpler. It should only be asymmetric between the two wells in order to introduce the discussed dynamics arising from {\it different} loss rates in the wells. \red{The quantitative decay, however, depends on the type of loss, and in particular we observe strong deviations from exponential decay for nonlinear two- and three-body loss. The latter might be relevant for probing different decay channels in experiments.}

The single-body case, i.e., with no interactions and single-body loss, can be solved analytically, even for an arbitrary long chain of wells, please see appendix A. The general case has been treated here in mean-field and one order beyond-mean field approximations.  Our BBR \red{beyond mean-field} approximation, here extended to include many-body loss, \red{proves reliable} for particle numbers which lie inbetween the small numbers for which exact simulations are still efficient and very large numbers for which mean-field is good enough, in particular also for longer chains, see our general derivation in appendix \ref{sec-appendix-b}. 

Future work may apply \red{the reported possibilities of dynamically controlling the out-coupled populations} to design \red{setups based on ultracold bosons for quantum synchronisation studies \cite{Witthaut2017} or} for atomtronic transport induced by a local leak \cite{Kordas2015, PRL2015, PhysRevLett.116.235302, PhysRevA.96.011601}.

\appendix

\section{Noninteracting case with single-body loss}
\label{sec-appendix-a}

In the noninteracting limit,  Eq. (\ref{eq:4}) corresponding to an $M$-site Bose-Hubbard chain in the presence of single-body losses can be put into the form $\partial_t \vec{\sigma}(t)=\mathbf{A}\vec{\sigma}(t)$. The solution becomes trivial when expressed in the basis of the matrix $\mathbf{A}$, for any set of initial conditions, the general solution is then given by $\vec{\sigma}(t)=\mathbf{U}(t)\vec{\sigma}(0)$, with:
\begin{equation}
\mathbf{U}(t)\equiv \mathbf{T}\exp(\mathbf{D}t)\mathbf{T^{-1}}=\mathbf{T}\,\textrm{diag}(\{e^{\lambda_i t}\})\mathbf{T^{-1}}
\label{eq:A1}
\end{equation}
Where $\mathbf{D}=\textrm{diag}(\{\lambda_i\})$, $\lambda_i$ denotes the $i$-th eigenvalue of $\mathbf{A}$ and $\mathbf{T}$ is the matrix formed from the corresponding eigenvectors that operates the change of basis.

Rewriting Eq. (\ref{eq:8}) in this way makes appear explicitly the eigenfrequencies that drive the system's dynamics and allows one to directly obtain a set of combinations of the SPDM's matrix elements ($\{(\mathbf{T^{-1}}\vec{\sigma})_i\}$) for each of which the time evolution is fully characterised by a single frequency ($\{\lambda_i\}$). 

For instance, in the two-well case with single-body loss occurring at the first well, $\vec{\sigma}$ and $\mathbf{A}$ can be expressed as follows:
\begin{equation}
\vec{\sigma}=\begin{pmatrix}\sigma_{11} \\ \sigma_{12} \\ \sigma_{21} \\ \sigma_{22} \end{pmatrix}\,;
\hspace{0.5cm}
\mathbf{A}=-J \begin{pmatrix}
x & -i & i & 0 \\ 
-i & x/2 & 0 & i \\ 
i & 0 & x/2 & -i \\ 
0 & i & -i & 0
\end{pmatrix}\,
\label{eq:A2}
\end{equation}
With the spectrum $\textrm{Sp}(\mathbf{A}/J)=\{-x/2+2y,-x/2-2y,-x/2,-x/2\}$. $\mathbf{D}$ is then given by:
\begin{equation}
\mathbf{D} =-\frac{\gamma_1}{2}\,\mathbf{I}_4+2J y\,\textrm{diag}(1,-1,0,0)
\label{eq:A3}
\end{equation}
The first term corresponds to an uniform damping that can be factorized out of (\ref{eq:A1}). The second term exhibits two different regimes depending on whether $y$ is real or imaginary. For a high enough rate of dissipation ($\gamma_1/2>2J$), the $\pm2J y$ eigenvalues are real and correspond to a decaying evolution of the populations $n_1 \equiv \sigma_{11}$ and $n_2 \equiv \sigma_{22}$; conversely, for a low rate of dissipation ($\gamma_1/2<2J$), they become imaginary and the solution presents oscillations in the SPDM's matrix elements. Finally, the second term vanishes for $\gamma_1/2=2J$ leading to an uniform exponential decay $\vec{\sigma}=\exp(-\gamma_1 t/2)\vec{\sigma}(0)$. Physically, this behaviour corresponds to the dissipation-induced lowering of the frequency of the oscillations of the populations, encoded in $y$, from the non-dissipative case $\omega=2J$ down to the non-oscillatory regime.

A mere examination of the eigenvalues of $\mathbf{A}$ thus suffices to identify the characteristic frequencies of the system evolution and to distinguish its different regimes for any Bose-Hubbard chain in the non-interacting limit. This provide a simple way to scrutinize the behaviour of longer chains.

In the case of an $M$-site BH chain, the frequencies of the non-dissipative case that will be lowered by the dissipation are given by the set $\{2J(\lambda_j-\lambda_{n-j})\}_j$, with $n\in[1,M]$ and $\lambda_\ell=-2\cos(\pi\ell/(M+1))$, as can be shown by diagonalising the Heisenberg equation satisfied by $\vec{a}\equiv (\hat{a}_1(t),\ldots,\hat{a}_M(t))^T$ and expressing explicitly the SPDM as the dyadic product $\sigma_{j,k}=\langle(\vec{a}\,^\dagger(t) \otimes \vec{a}(t))_{j,k}\rangle$. For periodic boundary conditions, solving the Heisenberg equation satisfied by $\vec{a}$ reduces to diagonalising a circulant matrix, and the $\mathbf{T}$ matrix can be given for any arbitrary length of the BH chain, leading to the same expression for the set of eigenfrequencies but with instead $\lambda_\ell=2\cos(2\pi\ell/M)$.

In our example of a two-well system, $\mathbf{T}$ is subsequently obtained from the eigenvectors of $\mathbf{A}$:
\begin{equation}
\mathbf{T}=\begin{pmatrix}\begin{smallmatrix}
1 & 1 & 1 & 0 \\ 
-i(y+x/4) & i(y+x/4) & 0 & 1 \\ 
i(y+x/4) & -i(y-x/4) & ix/2 & 1 \\ 
\tfrac{x(y-x/4)}{2}-1 & 1-\tfrac{x(y-x/4)}{2} & 1 & 0
\end{smallmatrix}\end{pmatrix}
\label{eq:A4}
\end{equation}
Finally, the analytical expression for the symmetric $\mathbf{U}(t)$ reads:
\begin{equation}
\begingroup
\renewcommand*{\arraystretch}{1.}
\mathbf{U}=\tfrac{e^{-\gamma_1 t/2}}{4y^2}\begin{pmatrix}
a_{-} & ib_{+}    	& -ib_{+}   & c \\
 . & 4y^2-c         & c     & -ib_{-} \\
 . & . & 4y^2-c & ib_{-} \\
.  &         .      &   .    & a_{+}
\end{pmatrix}
\endgroup
\label{eq:A5}
\end{equation}
Here we used the definitions
\begin{align}
a_\pm&=-2+((x/4)^2-2)\cosh(2J yt)\pm xy\sinh(2J y t)\\
b_\pm&=\pm\frac{x}{2}(1-\cosh(2Jyt))+2y\sinh(2J y t)\\
c&=2(\cosh(2J yt)-1) \, .
\end{align}
All the SPDM matrix elements can then be obtained from $\vec{\sigma}(t)=\mathbf{U}(t)\vec{\sigma}(0)$.


\section{BBR approximation for $\alpha$-body loss}
\label{sec-appendix-b}

Contrary to the mean-field approximation, in the Bogoliubov Back-reaction (BBR) approximation variances are kept in Eq. \eqref{eq:4} while higher moments are truncated as follows \cite{Anglin2001,Tikhonenkov2007}:
\begin{align}
\langle\hat{a}_j^\dagger\hat{a}_m^{\mathstrut}\hat{a}_k^\dagger\hat{a}_n^{\mathstrut}\hat{a}_r^\dagger\hat{a}_s^{\mathstrut}\rangle \bbrapprox& \langle \hat{a}_j^\dagger\hat{a}_m^{\mathstrut}\hat{a}_k^\dagger\hat{a}_n^{\mathstrut} \rangle\langle \hat{a}_r^\dagger\hat{a}_s^{\mathstrut} \rangle\nonumber\\
+ &\langle \hat{a}_j^\dagger\hat{a}_m^{\mathstrut}\hat{a}_r^\dagger\hat{a}_s^{\mathstrut} \rangle\langle \hat{a}_k^\dagger\hat{a}_n^{\mathstrut} \rangle+\langle \hat{a}_k^\dagger\hat{a}_n^{\mathstrut}\hat{a}_r^\dagger\hat{a}_s^{\mathstrut} \rangle\langle \hat{a}_j^\dagger\hat{a}_m^{\mathstrut} \rangle\nonumber\\
- &2\langle \hat{a}_j^\dagger\hat{a}_m^{\mathstrut} \rangle\langle \hat{a}_k^\dagger\hat{a}_n^{\mathstrut} \rangle\langle \hat{a}_r^\dagger\hat{a}_s^{\mathstrut} \rangle 
\label{eq:b1}
\end{align}
\begin{align}
\langle \hat{a}_j^\dagger\hat{a}_m^{\mathstrut}\hat{a}_k^\dagger\hat{a}_n^{\mathstrut}\hat{a}_r^\dagger\hat{a}_s^{\mathstrut}\hat{a}_a^\dagger\hat{a}_b^{\mathstrut} \rangle \,\bbrapprox\,& \langle \hat{a}_j^\dagger\hat{a}_m^{\mathstrut}\hat{a}_k^\dagger\hat{a}_n^{\mathstrut} \rangle\langle \hat{a}_r^\dagger\hat{a}_s^{\mathstrut}\hat{a}_a^\dagger\hat{a}_b^{\mathstrut} \rangle\nonumber\\
&\hspace{-80pt}+ \langle \hat{a}_j^\dagger\hat{a}_m^{\mathstrut}\hat{a}_r^\dagger\hat{a}_s^{\mathstrut} \rangle\langle \hat{a}_k^\dagger\hat{a}_n^{\mathstrut}\hat{a}_a^\dagger\hat{a}_b^{\mathstrut} \rangle + \langle \hat{a}_j^\dagger\hat{a}_m^{\mathstrut}\hat{a}_a^\dagger\hat{a}_b^{\mathstrut} \rangle\langle \hat{a}_k^\dagger\hat{a}_n^{\mathstrut}\hat{a}_r^\dagger\hat{a}_s^{\mathstrut} \rangle\nonumber\\
&\hspace{-80pt}-2\langle \hat{a}_j^\dagger\hat{a}_m^{\mathstrut} \rangle\langle \hat{a}_k^\dagger\hat{a}_n^{\mathstrut} \rangle\langle \hat{a}_r^\dagger\hat{a}_s^{\mathstrut} \rangle\langle \hat{a}_a^\dagger\hat{a}_b^{\mathstrut} \rangle \,.
\label{eq:b2}
\end{align}

This approximation takes approximately into account two-point correlations with a relative error of $1/N^2$. It is thus only valid for close to pure BEC states and large populations in all wells.

In the BBR approximation, the equations of motion of an $M$-site BH chain read
\begin{align}
i\frac{d}{dt}\sigma_{jk} = &-J\big(\sigma_{j,k+1}+\sigma_{j,k-1}-\sigma_{j+1,k}-\sigma_{j-1,k}\big)\nonumber\\
&+U\big(\Delta_{jkkk}+\sigma_{jk}\sigma_{kk}-\Delta_{jjjk}-\sigma_{jj}\sigma_{jk}\big)\nonumber\\
&-i\sum_{\ell=1}^M\frac{\gamma_\ell^{(\alpha)}}{2}S_{jk;\ell}^{(\alpha)}
\label{eq:b3}
\end{align}
\begin{align}
i\frac{d}{dt}\Delta_{jmkn}& =\nonumber\\ -J\big(&+\Delta_{j,m,k,n+1}+\Delta_{j,m,k,n-1}+\Delta_{j,m+1,k,n}\nonumber\\
&+\Delta_{j,m-1,k,n}-\Delta_{j,m,k+1,n}-\Delta_{j,m,k-1,n}\nonumber\\
&-\Delta_{j+1,m,k,n}-\Delta_{j-1,m,k,n}\big)\nonumber\\
+U\big(&+\sigma_{jm}(\Delta_{mmkn}-\Delta_{jjkn})\nonumber\\
&+\sigma_{kn}(\Delta_{jmnn}-\Delta_{jmkk})\nonumber\\
&+\Delta_{jmkn}(-\sigma_{jj}+\sigma_{mm}-\sigma_{kk}+\sigma_{nn})\big)\nonumber\\
&\hspace{-22pt}-i\sum_{\ell=1}^M\frac{\gamma_\ell^{(\alpha)}}{2}D_{jmkn;\ell}^{(\alpha)} \,.
\label{eq:b4}
\end{align}
Here the covariances are defined as \smash{$\Delta_{jmkn} \equiv \langle \hat{a}_j^\dagger\hat{a}_m^{\mathstrut}\hat{a}_k^\dagger\hat{a}_n^{\mathstrut} \rangle - \langle \hat{a}_j^\dagger\hat{a}_m^{\mathstrut} \rangle\langle \hat{a}_k^\dagger\hat{a}_n^{\mathstrut} \rangle$} and the matrices $S^{(\alpha)}$ and $D^{(\alpha)}$ denote the $\alpha$-body loss dissipative terms. These are defined as follows:
\begin{equation}
S_{jk;\ell}^{(\alpha)}=\big\langle \big[\hat{a}^{\dagger}_j\hat{a}^{\mathstrut}_k,\hat{a}_\ell^{\dagger\alpha}\big]\hat{a}_\ell^{\mathstrut\alpha} \big\rangle+\big\langle \hat{a}_\ell^{\dagger\alpha}\big[\hat{a}_\ell^{\mathstrut\alpha}, \hat{a}^{\dagger}_j\hat{a}^{\mathstrut}_k\big]\big\rangle \,.
\label{eq:b5}
\end{equation}
\begin{align}
&D_{jmkn;\ell}^{(\alpha)}=\nonumber\\
&\quad+\big\langle \big[\hat{a}^{\dagger}_j\hat{a}^{\mathstrut}_m\hat{a}^{\dagger}_k\hat{a}^{\mathstrut}_n,\hat{a}_\ell^{\dagger\alpha}\big]\hat{a}_\ell^{\mathstrut\alpha} \big\rangle+\big\langle \hat{a}_\ell^{\dagger\alpha}\big[\hat{a}_\ell^{\mathstrut\alpha}, \hat{a}^{\dagger}_j\hat{a}^{\mathstrut}_m\hat{a}^{\dagger}_k\hat{a}^{\mathstrut}_n,\big]\big\rangle\nonumber\\
&\quad-\sigma_{j,m}S_{kn;\ell}^{(\alpha)}-\sigma_{k,n}S_{jm;\ell}^{(\alpha)} \,.
\label{eq:b6}
\end{align}
The commutators in above formulae can be identified from the traced $\alpha$-body Liouvillian, as defined in Eq. (\ref{eq:5}), and the second line of Eq. (\ref{eq:b6}) results from deriving the second term of the covariance definition. The explicit expressions of $S^{(\alpha)}$ and $D^{(\alpha)}$ are given hereafter.

\subsection{Single-body losses ($\alpha=1$)}

For single-body loss, the calculation is straightforward and gives
\begin{align}
S_{jk;\ell}^{(1)}=&(\delta_{j,\ell}+\delta_{k,\ell})\sigma_{jk}\label{eq:b7}\\
D_{jmkn;\ell}^{(1)}=&(\delta_{j,\ell}+\delta_{m,\ell}+\delta_{k,\ell}+\delta_{n,\ell})\Delta_{jmkn;\ell}\nonumber\\
&-2\delta_{m,\ell}\delta_{k,\ell}\sigma_{jn} \,.
\label{eq:b8}
\end{align}
Which does not require any further truncation.

\subsection{Two-body losses ($\alpha=2$)}

For two-body loss, Eq. (\ref{eq:b5}) yields four-point correlation functions without the need of any truncation:
\begin{equation}
S_{jk;\ell}^{(2)}=(\delta_{j,\ell}+ \delta_{k,\ell})\big(2\Delta_{j\ell\ell k}+\sigma_{jk}f_2(n_\ell)\big) \,.
\end{equation}
Two-body loss, alike interactions, thus induces a coupling of the equations of motion of $n$-point correlation functions and $(n+1)$-point correlation functions.

The additional difficulty of two-body loss comes from Eq. (\ref{eq:b6}). Indeed, moments composed of strings of six operators arise from the commutators. Prior to performing their BBR truncation, these have to be rewritten into traces of products of three density-like operators, i.\,e., alternating creation and annihilation operators. The expression obtained after truncating six-point correlation functions depends on the reordering of the operators' indices, which is not unique. However, these different possible expressions only differ by subdominant lower moments in the mean-field limit. We choose an order such that the indices appear in the order of those of the covariance, that is $jmkn$, which yields
\begin{align}
\smash{D_{jmkn;\ell}^{(2)} =}\hspace{-20pt}&\nonumber\\
&2(\delta_{j,\ell}+\delta_{k,\ell})\big(\langle\hat{a}_\ell^\dagger\hat{a}_\ell^{\mathstrut}\hat{a}_j^\dagger\hat{a}_m^{\mathstrut}\hat{a}_k^\dagger\hat{a}_n^{\mathstrut}\rangle-\langle\hat{a}_j^\dagger\hat{a}_m^{\mathstrut}\hat{a}_k^\dagger\hat{a}_n^{\mathstrut}\rangle\big)\nonumber\\
+&2(\delta_{m,\ell}+\delta_{n,\ell})\big(\langle\hat{a}_j^\dagger\hat{a}_m^{\mathstrut}\hat{a}_k^\dagger\hat{a}_n^{\mathstrut}\hat{a}_\ell^\dagger\hat{a}_\ell^{\mathstrut}\rangle-\langle\hat{a}_j^\dagger\hat{a}_m^{\mathstrut}\hat{a}_k^\dagger\hat{a}_n^{\mathstrut}\rangle\big)\nonumber\\
-&2(2\delta_{m,\ell}\delta_{k,\ell}+\delta_{j,\ell}\delta_{k,\ell}+\delta_{m,\ell}\delta_{n,\ell})\langle\hat{a}_j^\dagger\hat{a}_m^{\mathstrut}\hat{a}_k^\dagger\hat{a}_n^{\mathstrut}\rangle\nonumber\\
+&2\times 4\delta_{m\ell}\delta_{k\ell}\sigma_{jn}\nonumber\\
-&2(\delta_{j,\ell}+\delta_{m,\ell})\Delta_{j\ell\ell m}\sigma_{kn}+2(\delta_{k,\ell}+\delta_{n,\ell})\sigma_{jm}\Delta_{k\ell\ell n}\nonumber\\
-&2(\delta_{j,\ell}+\delta_{m,\ell}+\delta_{k,\ell}+ \delta_{n,\ell})\sigma_{jm}\sigma_{kn}f_2(n_\ell)/2 \,.
\label{eq:b10}
\end{align}
This ordering ensures that after truncating, the leading moment ($\sim N^3$) cancels the last line of Eq. (\ref{eq:b10}) in the mean-field limit.

Eq. (\ref{eq:b10}) is then truncated as in Eq. (\ref{eq:b1}), leading to the final expression
\begin{align}
\smash{D_{jmkn;\ell}^{(2)} =}\hspace{-20pt}&\nonumber\\
&(\delta_{j,\ell}+\delta_{m,\ell}+\delta_{k,\ell}+\delta_{n,\ell})(n_\ell-1)\Delta_{jmkn}\nonumber\\
+&(\delta_{j,\ell}\Delta_{\ell\ell kn}+\delta_{m,\ell}\Delta_{kn\ell\ell})\sigma_{jm}\nonumber\\
+&(\delta_{k,\ell}\Delta_{\ell\ell jm}+\delta_{n,\ell}\Delta_{jm\ell\ell})\sigma_{kn}\nonumber\\
-& (2\delta_{m,\ell}\delta_{k,\ell}+\delta_{j,\ell}\delta_{k,\ell}+\delta_{m,\ell}\delta_{n,\ell})(\Delta_{jmkn}+\sigma_{jm}\sigma_{kn})\nonumber\\
+& 4\delta_{m\ell}\delta_{k\ell}\sigma_{jn} \,.
\label{eq:b11}
\end{align}

\subsection{Three-body losses ($\alpha=3$)}

Three-body loss involves the additional difficulty of having to perform the BBR truncation in the contribution to the SPDM's equation of motion as well, leading to
\begin{align}
S_{jk;\ell}^{(3)}=\smash{\sum_{\ell=1}^M\frac{\gamma_\ell^{(3)}}{2}(\delta_{j,\ell}+ \delta_{k,\ell})}\big(&3\Delta_{j\ell\ell k}(2n_\ell-3)\nonumber\\
&+\sigma_{jk}(3\Delta_{\ell\ell \ell\ell}+f_3(n_\ell))\big) \,.
\end{align}
Three-body loss thus induces an additional coupling of the equations of motion between the $n$-point correlation functions and the $(n+2)$-point correlation functions.

In this case, the reordering involves straightforward but lengthy calculations leading to the following expression
\begin{align}
\smash{D_{jmkn;\ell}^{(3)}/3 =}\hspace{-50pt}&\nonumber\\
&\hspace{-4pt}(\delta_{m,\ell}+\delta_{n,\ell})\big(\langle\hat{a}_j^\dagger\hat{a}_m^{\mathstrut}\hat{a}_k^\dagger\hat{a}_n^{\mathstrut}\hat{a}_\ell^\dagger\hat{a}_\ell^{\mathstrut}\hat{a}_\ell^\dagger\hat{a}_\ell^{\mathstrut}\rangle-3\langle\hat{a}_j^\dagger\hat{a}_m^{\mathstrut}\hat{a}_k^\dagger\hat{a}_n^{\mathstrut}\hat{a}_\ell^\dagger\hat{a}_\ell^{\mathstrut}\rangle\nonumber\\
&\hspace{155pt}+2\langle\hat{a}_j^\dagger\hat{a}_m^{\mathstrut}\hat{a}_k^\dagger\hat{a}_n^{\mathstrut}\rangle\big)\nonumber\\
+&(\delta_{j,\ell}+\delta_{k,\ell})\big(\langle\hat{a}_\ell^\dagger\hat{a}_\ell^{\mathstrut}\hat{a}_\ell^\dagger\hat{a}_\ell^{\mathstrut}\hat{a}_j^\dagger\hat{a}_m^{\mathstrut}\hat{a}_k^\dagger\hat{a}_n^{\mathstrut}\rangle-3\langle\hat{a}_\ell^\dagger\hat{a}_\ell^{\mathstrut}\hat{a}_j^\dagger\hat{a}_m^{\mathstrut}\hat{a}_k^\dagger\hat{a}_n^{\mathstrut}\rangle\nonumber\\
&\hspace{155pt}+2\langle\hat{a}_j^\dagger\hat{a}_m^{\mathstrut}\hat{a}_k^\dagger\hat{a}_n^{\mathstrut}\rangle\big)\nonumber\\
-&(\delta_{j,\ell}+\delta_{m,\ell})\sigma_{kn}\Delta_{j\ell\ell m}(2n_\ell-3)\nonumber\\
-&(\delta_{k,\ell}+\delta_{n,\ell})\sigma_{jm}\Delta_{k\ell\ell n}(2n_\ell-3)\nonumber\\
-&(\delta_{j,\ell}+\delta_{m,\ell}+\delta_{k,\ell}+ \delta_{n,\ell})\sigma_{jm}\sigma_{kn}(\Delta_{\ell\ell\ell\ell}+f_3(n_\ell)/3)\nonumber\\
+&(8\delta_{m,\ell}\delta_{k,\ell}+4\delta_{j,\ell}\delta_{k,\ell}+4\delta_{m,\ell}\delta_{n,\ell}+(\delta_{j,\ell}+\delta_{n,\ell})\delta_{m,\ell}\delta_{k,\ell})\nonumber\\
&\hspace{173pt}\times\langle\hat{a}_j^\dagger\hat{a}_m^{\mathstrut}\hat{a}_k^\dagger\hat{a}_n^{\mathstrut}\rangle\nonumber\\
-&12\delta_{m,\ell}\delta_{k,\ell}\langle\hat{a}_j^\dagger\hat{a}_n^{\mathstrut}\rangle \,.
\label{eq:b13}
\end{align}
Once again, this specific reordering of indices ensures that, in Eq. (\ref{eq:b13}), the leading term ($\sim N^4$) that would emerge from the BBR truncation of the two first lines would cancel the antepenultimate line in the mean-field limit.

In terms of the SPDM and the covariances, Eq. (\ref{eq:b13}) can be rewritten in the following way
\begin{align}
\smash{D_{jmkn;\ell}^{(3)}/3 =}\hspace{-50pt}&\nonumber\\
&(\delta_{j,\ell}+\delta_{m,\ell}+\delta_{k,\ell}+ \delta_{n,\ell})\Delta_{jmkn}(\Delta_{\ell\ell \ell\ell}+n_\ell^2-3n_\ell+2)\nonumber\\
+&2(\delta_{j,\ell}+\delta_{k,\ell})\Delta_{\ell\ell jm}\Delta_{\ell\ell kn}+2(\delta_{m,\ell}+\delta_{n,\ell})\Delta_{jm\ell\ell}\Delta_{kn\ell\ell}\nonumber\\
+&(2n_\ell-3)\sigma_{jm}(\delta_{j,\ell}\Delta_{\ell\ell kn}+\delta_{m,\ell}\Delta_{kn\ell\ell})\nonumber\\
+&(2n_\ell-3)\sigma_{kn}(\delta_{k,\ell}\Delta_{\ell\ell jm}+\delta_{n,\ell}\Delta_{jm\ell\ell})\nonumber\\
+&2\big(\delta_{m,\ell}\delta_{k,\ell}(4-n_\ell)+\delta_{j,\ell}\delta_{k,\ell}(2-n_\ell)+\delta_{m,\ell}\delta_{n,\ell}(2-n_\ell)\big)\nonumber\\
&\hspace{145pt}\times(\Delta_{jmkn}+\sigma_{jm}\sigma_{kn})\nonumber\\
+&(\delta_{j,\ell}+\delta_{n,\ell})\delta_{m,\ell}\delta_{k,\ell}(\Delta_{jmkn}+\sigma_{jm}\sigma_{kn})-12\delta_{m,\ell}\delta_{k,\ell}\sigma_{jn}\nonumber\\
-&(\delta_{m,\ell}\delta_{k,\ell}+2\delta_{m,\ell}\delta_{n,\ell})(\Delta_{jm\ell\ell}\sigma_{kn}+\Delta_{kn\ell\ell}\sigma_{jm})\nonumber\\
-&(\delta_{m,\ell}\delta_{k,\ell}+2\delta_{j,\ell}\delta_{k,\ell})(\Delta_{\ell\ell jm}\sigma_{kn}+\Delta_{\ell\ell kn}\sigma_{jm}) \,.
\label{eq:b14}
\end{align}

\subsection{Validity and stability of the BBR method}
\label{sec:B4}

The relative error committed in the covariances by the BBR approximation scales as $1/N^2$. Moreover, the state of the system has to remain close to a pure BEC state so that the dominant quadratic dependence on the fields ensures that the moments can be truncated as gaussian variables, which amounts to consider close to mean-field situations. Therefore, it becomes exact in the limit $N\rightarrow+\infty$ while keeping $UN/J$ constant, and it naturally converges to mean-field in this limit. As an indicator, the value of the covariances should remain close to zero, or equivalently, the $g^{(2)}_{j,k}$ matrix elements should remain close to one; please see the discussion in sect. \ref{sec-5} around Fig. \ref{fig:8}.

Figures \ref{fig:5} and \ref{fig:6} show that the BBR approximation performs well even down to $N\sim 10$ for $UN_0/J=5$ for two and three-body loss in the Josephson oscillations regime. Moreover, strong dissipation is known \cite{Witthaut2011} to increase the precision of this method by suppressing density-density correlations and thus making the covariances decline. However, in the case of self-trapping in the presence of two or three-body loss, if one of the populations becomes too low (in particular less than $\sim\alpha$), the BBR approximation leads to numerically unstable behaviour.

\begin{acknowledgments}
We thank Dirk Witthaut for a critical reading of the manuscript.
Z.D. acknowledges support by an internship grant from the Université 
Paris-Saclay for his stay at Parma. L.S. acknowledges for partial support 
the 2016 BIRD project "Superfluid properties of Fermi gases in 
optical potentials" of the University of Padova. 
\end{acknowledgments}

\vfill

%

\end{document}